\documentclass[aps,prb,twocolumn,preprintnumbers,citeautoscript]{revtex4-1}

\usepackage{amsmath,amsfonts,amssymb,bm,graphicx}
\usepackage{color}

\newcommand{\ket}[1]{\lvert #1\rangle}
\newcommand{\bra}[1]{\langle#1 \rvert}
\newcommand{\abs}[1]{\lvert #1 \rvert}

\newcommand{\bk}{\mathbf{k}}
\newcommand{\bq}{\mathbf{q}}

\begin{document}

\title{Flexural phonon scattering induced by electrostatic gating in graphene}

\author{Tue Gunst}
\email{Tue.Gunst@nanotech.dtu.dk}
\affiliation{Department of Micro- and Nanotechnology (DTU Nanotech), Center for
  Nanostructured Graphene (CNG), Technical University of Denmark, DK-2800
  Kgs. Lyngby, Denmark}
\author{Kristen Kaasbjerg}
\affiliation{Department of Micro- and Nanotechnology (DTU Nanotech), Center for
  Nanostructured Graphene (CNG), Technical University of Denmark, DK-2800
  Kgs. Lyngby, Denmark}
\author{Mads Brandbyge}
\affiliation{Department of Micro- and Nanotechnology (DTU Nanotech), Center for
  Nanostructured Graphene (CNG), Technical University of Denmark, DK-2800
  Kgs. Lyngby, Denmark}

\date{\today}

\begin{abstract}
Graphene has an extremely high carrier mobility partly due to its planar mirror symmetry inhibiting scattering by the highly occupied acoustic flexural phonons. Electrostatic gating of a graphene device can break the planar mirror symmetry yielding a coupling mechanism to the flexural phonons. We examine the effect of the gate-induced one-phonon scattering on the mobility for several gate geometries and dielectric environments using first-principles calculations based on density functional theory (DFT) and the Boltzmann equation. We demonstrate that this scattering mechanism can be a mobility-limiting factor, and show how the carrier density and temperature scaling of the mobility depends on the electrostatic environment. Our findings may explain the high deformation potential for in-plane acoustic phonons extracted from experiments and furthermore suggest a direct relation between device symmetry and resulting mobility.
\end{abstract}
\maketitle

The carrier mobility limited by electron-phonon (el-ph) scattering is an important performance indicator of emerging two-dimensional (2D) materials~\cite{geim_rise_2007}. Extremely high carrier mobilities have been measured for graphene which promises many exciting nanoelectronic and optoelectronic applications~\cite{geim_graphene:_2009}. However, the reported carrier mobilities vary significantly with atomic defects, grain boundaries, strain and charge impurities often being mentioned as the main mobility-limiting factors.\cite{geim_rise_2007,morozov_giant_2008,efetov_controlling_2010,chen_diffusive_2009,chen_ionic_2009,feldman_broken-symmetry_2009,castro_limits_2010,du_approaching_2008,dean_boron_2010,kim_synthesis_2012,couto_random_2014,buron_graphene_2015,wang_one-dimensional_2013} At the same time, the mobility of high-quality devices is approaching the intrinsic phonon-limited value~\cite{efetov_controlling_2010,dean_boron_2010,wang_one-dimensional_2013}, which underlines the importance of controlling the el-ph interaction in graphene devices.

Recently, the intrinsic mobility of 2D materials with broken planar mirror reflection ($\sigma_h$) symmetry, such as, e.g., silicene and germanene, has been demonstrated to be very low~\cite{fischetti_mermin-wagner_2016,gunst_first-principles_2016}. The explanation is found in a strong coupling to the flexural-acoustic (ZA) membrane mode in combination with an exceedingly high occupation of this mode due to its quadratic dispersion and constant density of phonon states (DOS).\cite{lindsay_flexural_2010} In graphene, however, the standard linear el-ph coupling vanishes for the flexural phonon due to the preserved $\sigma_h$ symmetry, and scattering requires two-phonon processes via the quadratic coupling. In suspended graphene, this has been demonstrated to be important~\cite{mariani_flexural_2008,castro_limits_2010,mariani_temperature-dependent_2010,gornyi_conductivity_2012,kerner_bending_2012,ochoa_scattering_2012}. Most studies of phonon-limited mobilities in supported graphene~\cite{manes_symmetry-based_2007,stauber_electronic_2007,hwang_acoustic_2008,shishir_velocity_2009,shishir_intrinsic_2009,chen_diffusive_2009,kaasbjerg_unraveling_2012,park_electronphonon_2014,sohier_phonon-limited_2014}, however, neglect flexural phonon scattering altogether.

\begin{figure}[!tbp]
\centering
{\includegraphics[width=0.99\linewidth]{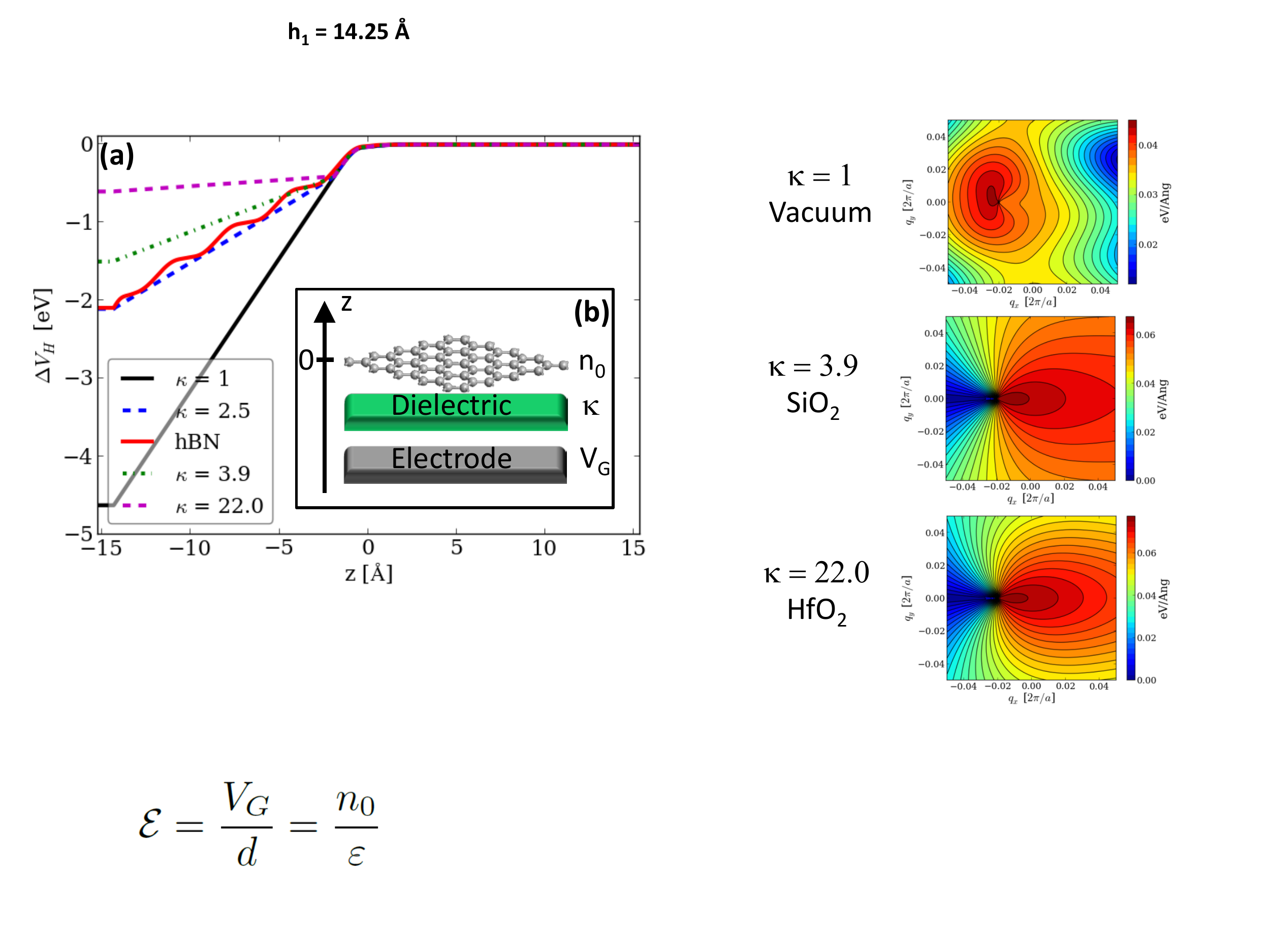}}\\
{\includegraphics[width=0.99\linewidth]{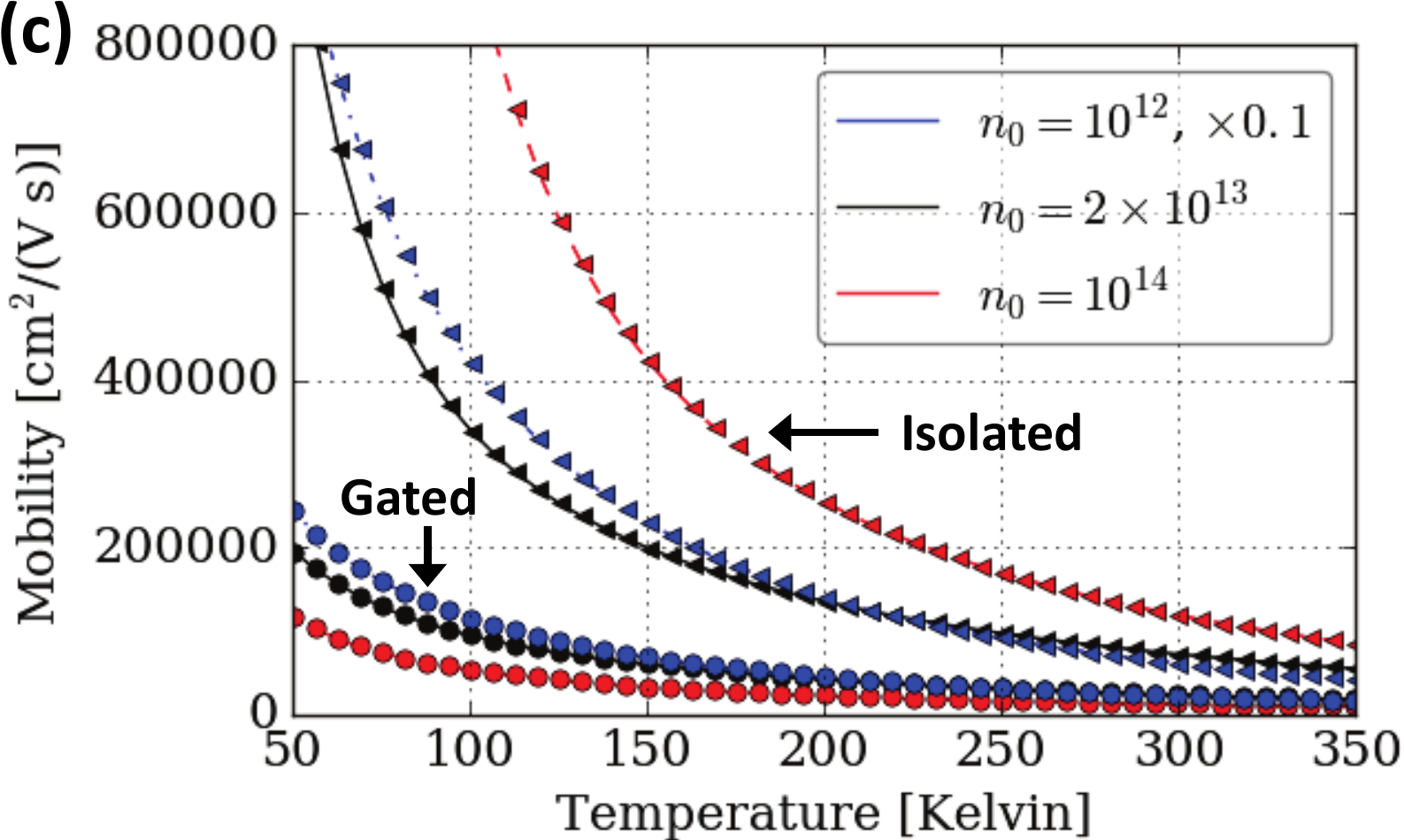}}
\caption{Broken $\sigma_h$ symmetry and mobility degradation in gated graphene devices. (a) Potential profile across the device in (b) for different dielectric substrates and a graphene carrier density of $n_0=2\times10^{13}$~cm$^{-2}$. The asymmetry in the potential across graphene triggers scattering with the out-of-plane flexural modes. (b) Device geometry with graphene separated from the gate electrode with voltage $V_G$ by a dielectric region with dielectric constant $\kappa$. (c) Mobility vs temperature at different carrier densities for a graphene device with $\kappa=2.5$ (circles: $\circ$) and isolated graphene with preserved $\sigma_h$ symmetry (triangles: $\triangleleft$); i.e., respectively, with and without field-induced flexural phonon scattering.}
\label{fig:SystemCapacitor}
\end{figure}
In this letter, we demonstrate that typical electrostatic gating and the dielectric environment in graphene devices can break the $\sigma_h$ symmetry and activate one-phonon scattering processes by flexural phonons. We show that this can lead to a degradation of the carrier mobility up to several orders of magnitude depending on the applied field and dielectric environment. The scattering from flexural phonons is particularly important in the low-temperature regime due to a much lower Bloch-Gr\"uneisen (BG) temperature. This leads to a temperature and density dependence of the mobility which differ from those for scattering off in-plane acoustic phonons~\cite{hwang_acoustic_2008,kaasbjerg_unraveling_2012}. We further discuss how field-induced scattering can be suppressed by employing a gate configuration that preserves the planar mirror symmetry of graphene.

The origin of the broken $\sigma_h$ symmetry in standard graphene device setups is illustrated in Figs.~\ref{fig:SystemCapacitor}(a) and~\ref{fig:SystemCapacitor}(b). A voltage $V_G$ is applied to the metallic gate electrode to control the carrier density $n_0$ in graphene. The electric field is efficiently screened by the doped graphene layer, and the potential drop along the $z$ direction occurs mainly on the side of the graphene layer facing the gate dielectric. This resembles the situation in a parallel-plate capacitor where the electric field is confined to the region between the plates and is given by
\begin{eqnarray}
  \label{eqn:Field}
  \mathcal{E} = \frac{V_{G}}{d} = \frac{e n_0}{\kappa\epsilon_0} ,
\end{eqnarray}
where $\kappa$ is the dielectric constant and $\epsilon_0$ is the vacuum permittivity. The efficient screening provided by the ``graphene plate'' hence results in a pronounced asymmetry in the potential profile across graphene. Like in carbon-nanotube mechanical resonators~\cite{lassagne_coupling_2009,steele_strong_2009,ganzhorn_dynamics_2012,benyamini_real-space_2014}, the motion of the graphene layer in this potential induces a finite coupling to the flexural phonon modes. We term this a 'field-induced' coupling, and as we here demonstrate, the strength of the field-induced coupling is governed by the gate potential and dielectric environment.

In practice, the potential in Fig.~\ref{fig:SystemCapacitor}(a) is obtained by solving the Poisson equation and electronic structure self-consistently at finite gate voltage and carrier density based on first-principles DFT simulations~\cite{FN_settings}. The el-ph coupling is calculated at each value of the gate voltage (or carrier density) in order to explicitly include the screened gate potential in the el-ph coupling~\cite{FN_PHandEPH}. We subsequently solve the Boltzmann equation in the relaxation-time approximation using the first-principles parameters as implemented in the Atomistix ToolKit~\cite{gunst_first-principles_2016,ATK}.
We only include the field-induced one-phonon ZA scattering.

We consider dielectric regions with $\kappa=1$ (vacuum), $\kappa=3.9$ (SiO$_2$) or $\kappa=22.0$ (HfO$_2$) corresponding to high-$\kappa$ dielectrics~\cite{fischetti_effective_2001}. To demonstrate the validity of a continuum description of the gate dielectric, we start by comparing the potential profile to that of graphene on top of four atomic layers of hexagonal boron nitride (h-BN). From Fig.~\ref{fig:SystemCapacitor}(a), we obtain a perfect match with a dielectric constant of h-BN of $\kappa \approx 2.5$ which is within the range of previously reported values of 2--4~\cite{kim_synthesis_2012}. We further extract the distance below graphene, $\sim 2.0$~\AA, at which the electric field is essentially unscreened. This distance is relevant for other layered materials with similar interlayer distances~\cite{hod_graphite_2012}.

In Fig.~\ref{fig:SystemCapacitor}(c), we demonstrate the effect of the broken $\sigma_h$ symmetry on the phonon-limited mobility
by including field-induced one-phonon ZA scattering, while we neglect two-phonon scattering. The figure shows the mobility as a function of temperature for three densities, $n_0=10^{12}$~cm$^{-2},\,2\times10^{13}$~cm$^{-2}$ and $10^{14}$~cm$^{-2}$, in the case of (i) an isolated graphene layer with preserved $\sigma_h$ symmetry (triangles), and (ii) the $\kappa=$2.5 device structure described above (circles). Examining the mobility at $T=300$~K, we find a reduction from approximately 606.000~$\mathrm{cm}^2/\mathrm{V}\,\mathrm{s}$ to 208.500~$\mathrm{cm}^2/\mathrm{V}\,\mathrm{s}$ at $n_0=10^{12}$~cm$^{-2}$ (blue; scaled by a factor $0.1$). For $n_0=2\times10^{13}$~cm$^{-2}$, the mobility is degraded from approximately 71.600~$\mathrm{cm}^2/\mathrm{V}\,\mathrm{s}$ to 22.400~$\mathrm{cm}^2/\mathrm{V}\,\mathrm{s}$ (black), and at $n_0=10^{14}$~cm$^{-2}$ the reduction is even more significant with a reduction from 118.000~$\mathrm{cm}^2/\mathrm{V}\,\mathrm{s}$ to 11.400~$\mathrm{cm}^2/\mathrm{V}\,\mathrm{s}$ (red). Overall, the mobility reduction is more pronounced in the low-temperature regime. Notice also that the mobility of isolated graphene increases with $n_0$ from $n_0=2\times10^{13}$~cm$^{-2}$ to $n_0=10^{14}$~cm$^{-2}$, while the opposite is observed for gated graphene. We will elaborate on this below.

The strong reduction of the mobility can be attributed to the field-induced el-ph coupling to the flexural modes. Even for suspended/free-hanging graphene ($\kappa=1$ model), we find that the field-induced coupling has detrimental impact on the transport~\cite{mariani_flexural_2008,castro_limits_2010,mariani_temperature-dependent_2010}.
For free-hanging graphene, the quadratic dispersion and constant DOS of the ZA mode gives rise to a very large phonon population in the long-wavelength limit. Therefore, even a coupling constant significantly smaller than those to the in-plane TA and LA modes~\cite{kaasbjerg_unraveling_2012,gunst_first-principles_2016} can lead to strong scattering off the flexural phonons. For graphene supported by a gate dielectric, the flexural modes are modified by the interaction with the substrate which result in (i) hybridization with the Rayleigh surface modes of the substrate, and (ii) a gap opening in the dispersion $\omega_q=\sqrt{b^2 q^4 + \omega_0^2}$, with the size of the gap, $\omega_0$, reflecting the substrate interaction strength~\cite{mariani_flexural_2008,mariani_temperature-dependent_2010,gornyi_conductivity_2012,ong_effect_2011,amorim_flexural_2013,slotman_phonons_2014}.
This introduces an effective cutoff $q_c=\sqrt{\omega_0/b}$ below which flexural phonon scattering is suppressed~\cite{castro_limits_2010}. For free-hanging graphene, an additional cutoff, $L = 2\pi/q_c$, related to a length scale $L$ set by defects or higher-order interaction effects is introduced. In our calculations, we account for the above-mentioned effects via a cutoff related to the computational phonon momentum resolution ($\bq$-mesh), and use a conservative $L=4$nm.
\begin{figure}[!tbp]
\centering
{\includegraphics[width=0.99\linewidth]{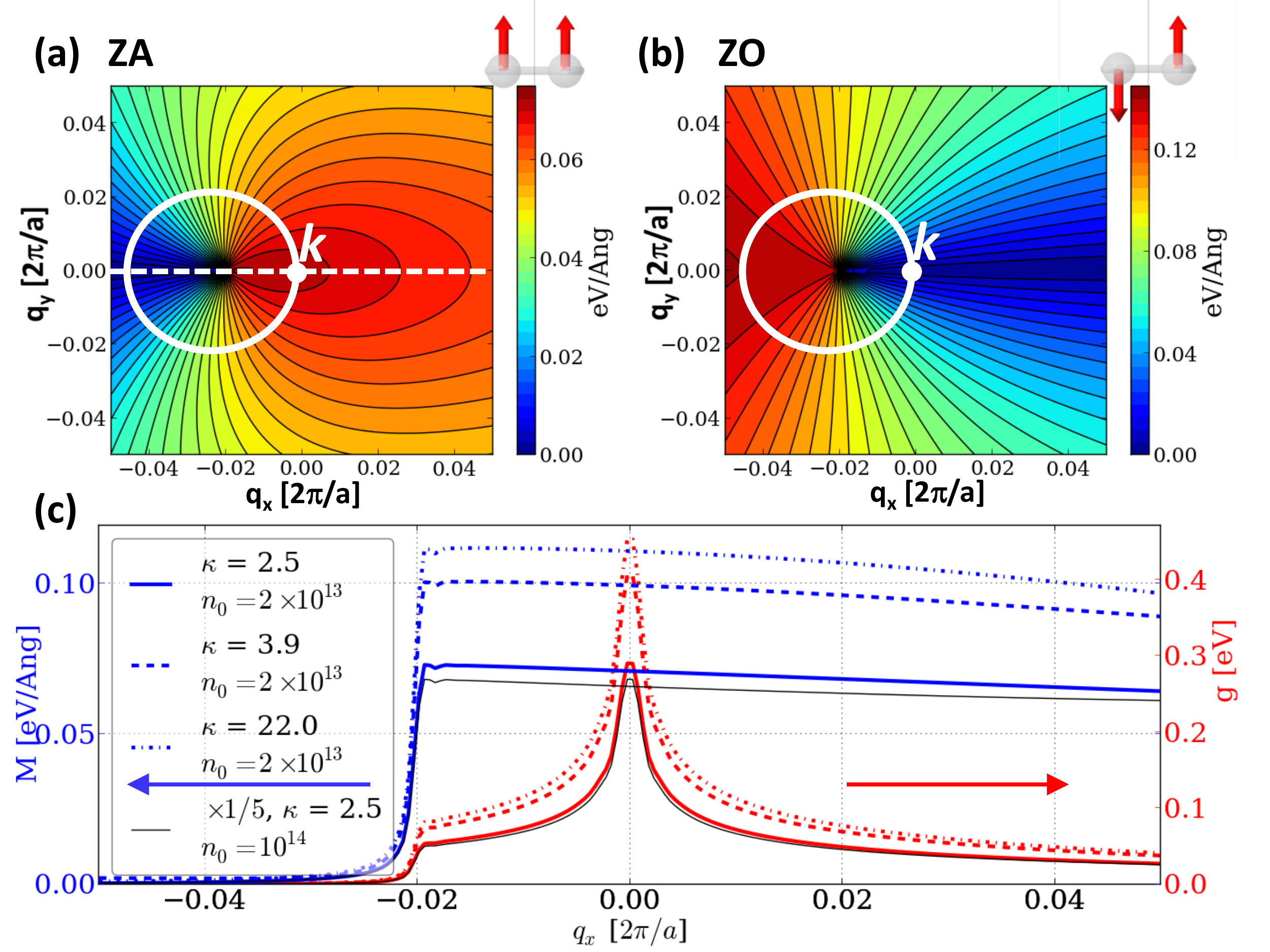}}
\caption{Field-induced el-ph couplings to flexural phonons. (a),(b) Coupling matrix element $M_{\bk\bk'}^\lambda$ for the (a) ZA and (b) ZO phonons at a carrier density of $n_0=2\times10^{13}$cm$^{-2}$ and $\kappa=2.5$. The $\mathbf{k}$ point is positioned $300$~meV above the Dirac point and the circles  indicate constant energy surfaces on the Dirac cones. (c) $q$ dependence of the coupling to the ZA phonon along the dashed line in (a) for different electrostatic conditions and carrier densitites.}
\label{fig:DOSratioAndDP}
\end{figure}

In Fig.~\ref{fig:DOSratioAndDP}(a) and \ref{fig:DOSratioAndDP}(b) we show the calculated el-ph couplings $g_{\bk\bk'}^\lambda = l_{\lambda q} M_{\bk\bk'}$ to the out-of-plane ZA and ZO modes, where $l_{\lambda q} = (\hbar/2M\omega_{\lambda q})^{1/2}$ is the characteristic length, $M$ is the unit cell mass, and $M_{\bk\bk'}^\lambda=\bra{\psi_{\bk'}} \delta V_{\lambda\bq}\ket{\psi_\bk}$, $\bq=\bk'-\bk$, is the coupling-matrix element between electronic states~\cite{gunst_first-principles_2016}. When the mirror reflection symmetry is broken, the out-of-plane atomic displacements associated with flexural phonons shift the energies of the $A$ and $B$ sublattice. In a low-energy description of the graphene Dirac cones, $H(\bk)=\hbar v_F \bk \cdot \bm{\sigma}$, the field-induced coupling to the ZA mode can be expressed as
\begin{equation}
  \label{eq:M}
  M_{\bk\bk'}^\text{ZA} = \bra{\chi_{\bk'}} \bm{M}^\text{ZA} \ket{\chi_{\bk}}
  , \quad
  \bm{M}^\text{ZA} =  D_0(V_G) \bm{\sigma}_0
\end{equation}
where $\ket{\chi_\bk}=[1,e^{i \phi_\bk}]^T/\sqrt{2}$ is the pseudospin eigenvector in the $A,B$ sublattice, $\bm{\sigma}_0$ is the identity matrix and $D_0(V_G)$ is the field-induced coupling constant. The field-induced coupling then becomes $\abs{M^\text{ZA}_{\bk\bk'}}^2 = D_0^2(V_G) \cos^2({\phi_{\bk\bk'}/2})$ where $\phi_{\bk\bk'}=\phi_\bk-\phi_{\bk'}$ is the scattering angle. Analogous to in-plane LA phonon scattering, this matrix element suppresses backscattering ($\phi_{\bk\bk'}=\pi$), which is consistent with the first-principles result in Fig.~\ref{fig:DOSratioAndDP}(a). This is reversed for the ZO mode in Fig.~\ref{fig:DOSratioAndDP}(b). Here $\mathbf{M}^\text{ZO} \propto \bm{\sigma}_z$ and hence forward scattering is suppressed. Likewise, the field-induced ZA coupling in Eq.~\eqref{eq:M} contrasts the situation for the intrinsic coupling to the ZA mode in nonplanar 2D materials like, e.g., silicene and germanene where $\mathbf{M} \propto \bm{\sigma}_z$, i.e., the sublattices energies are shifted in opposite directions~\cite{fischetti_mermin-wagner_2016}. This has important consequences since backscattering results in a more dramatic mobility reduction in such Dirac materials.

Contrary to in-plane acoustic phonons where the matrix element is linear in $q$, $M_{\bk\bk'}^\text{TA/LA} \propto q$, the ZA coupling in Eq.~\eqref{eq:M} is independent on $q$. As a consequence, $g_{\bk\bk'}^\text{ZA}$ increases drastically for $q\rightarrow q_c$ as demonstrated by our first-principles results in Fig.~\ref{fig:DOSratioAndDP}(c) which shows the $q$ dependence of the ZA coupling along the dashed line in Fig.~\ref{fig:DOSratioAndDP}(a) for different dielectric environments and carrier densities. Here, $M_{\bk\bk'}$ has a finite value as $q\rightarrow 0$ and is almost constant for $\phi_{\bk\bk'}=0$. By comparing the results at $n_0=2\times10^{13}$~cm$^{-2}$ and $n_0=10^{14}$~cm$^{-2}$ (the two full lines) we note that an increase in the carrier density [or gate voltage, cf. Eq.~\eqref{eqn:Field}] by a factor of five also increases the field-induced el-ph coupling by a factor of five, indicating that it is linearly dependent on the carrier density, $D_0(V_G) \sim n_0$.

In addition to the applied electric field, the dielectric environment can also affect the field-induced coupling to the flexural phonon. Often, dielectric engineering~\cite{jena_enhancement_2007,konar_effect_2010} utilizing high-$\kappa$ gate dielectrics or suspension in high-$\kappa$ liquids~\cite{newaz_probing_2012,shishir_intrinsic_2009,chen_ionic_2009} to screen out scattering from charged impurities is exploited to improve the carrier mobility. Such procedures will, however, not necessarily weaken the field-induced flexural-phonon scattering mechanism. On the contrary, Fig.~\ref{fig:DOSratioAndDP}(c) shows that an increase in the dielectric constant of the gate dielectric enhances the coupling constant from $D_0=0.07$~eV/{\AA} at $\kappa=2.5$ to $D_0=0.11$~eV/{\AA} at $\kappa=22$. However, compared to the field-induced coupling we find that this effect is of secondary importance. Defining $D_0(V_G) \equiv \gamma n_0$ we extract $\gamma=35$~$(55)$~eV$\cdot${\AA} for $\kappa=2.5$~$(22)$.

Having highlighted the fundamental properties of the field-induced flexural el-ph coupling, we now return to its impact on the mobility and discuss possible experimental verifications of the mechanism.
Already from Fig.~\ref{fig:SystemCapacitor}(c) it is evident that field-induced flexural phonon scattering gives rise to a different temperature and density dependence of the mobility. 

\begin{figure}[!tbp]
\centering
{\includegraphics[width=0.99\linewidth]{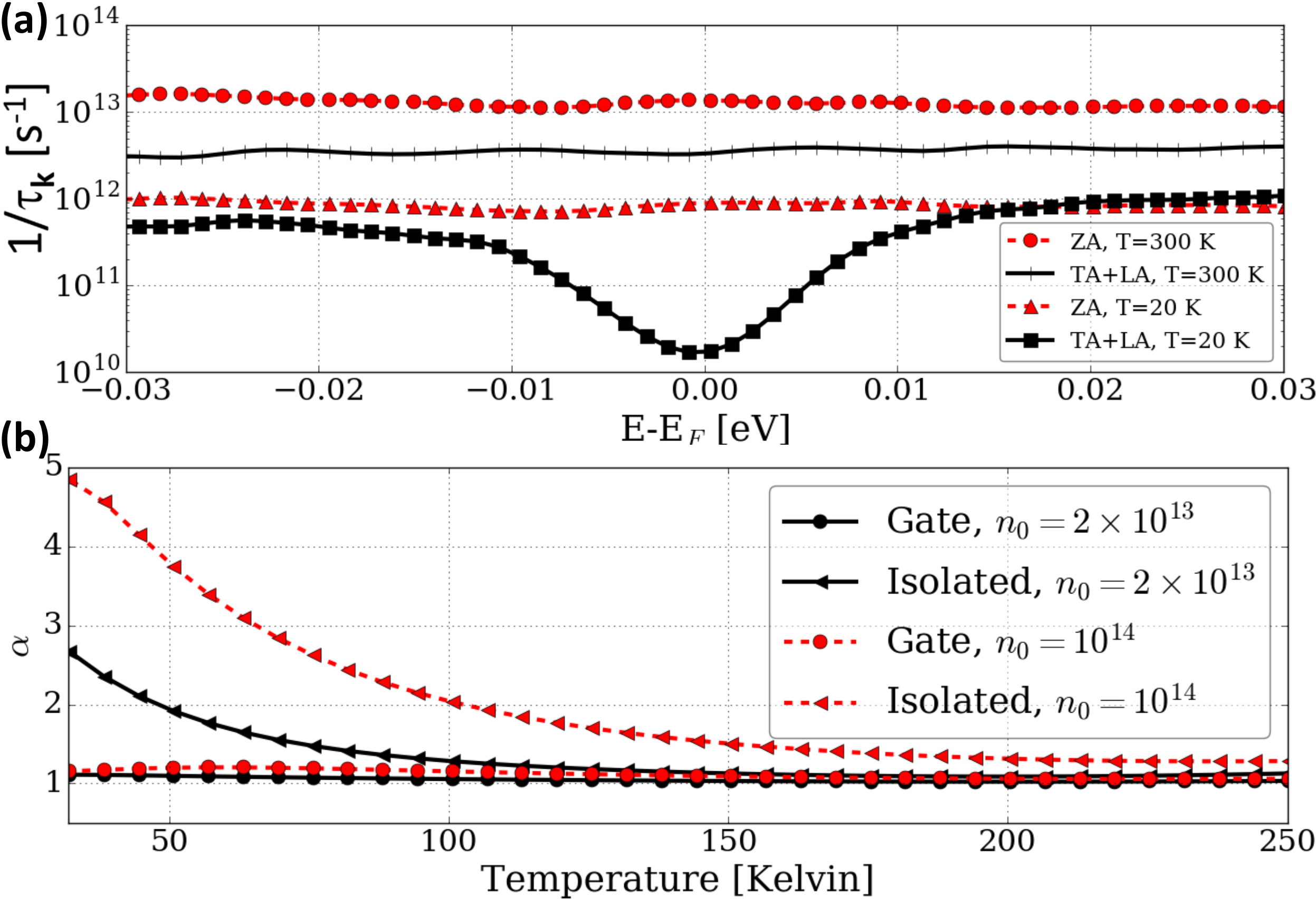}}
\caption{Temperature dependence of the field-induced ZA scattering limited mobility. (a) Scattering rate near the Fermi level for the in-plane TA+LA and the flexural ZA phonons at low and room temperature, $n_0=2\times10^{13}$cm$^{-2}$ and $\kappa=22$. (b) Temperature exponent, $\alpha = -d\mathrm{log}(\mu_e)/d\mathrm{log}(T)$, as a function of temperature for the mobilities in Fig.~\ref{fig:SystemCapacitor}(c). The field-induced ZA scattering dominates at room temperature and changes the temperature exponent in the low temperature regime.}
\label{fig:TemperatureDependence}
\end{figure}
\begin{figure}[t]
	\centering
	{\includegraphics[width=0.99\linewidth]{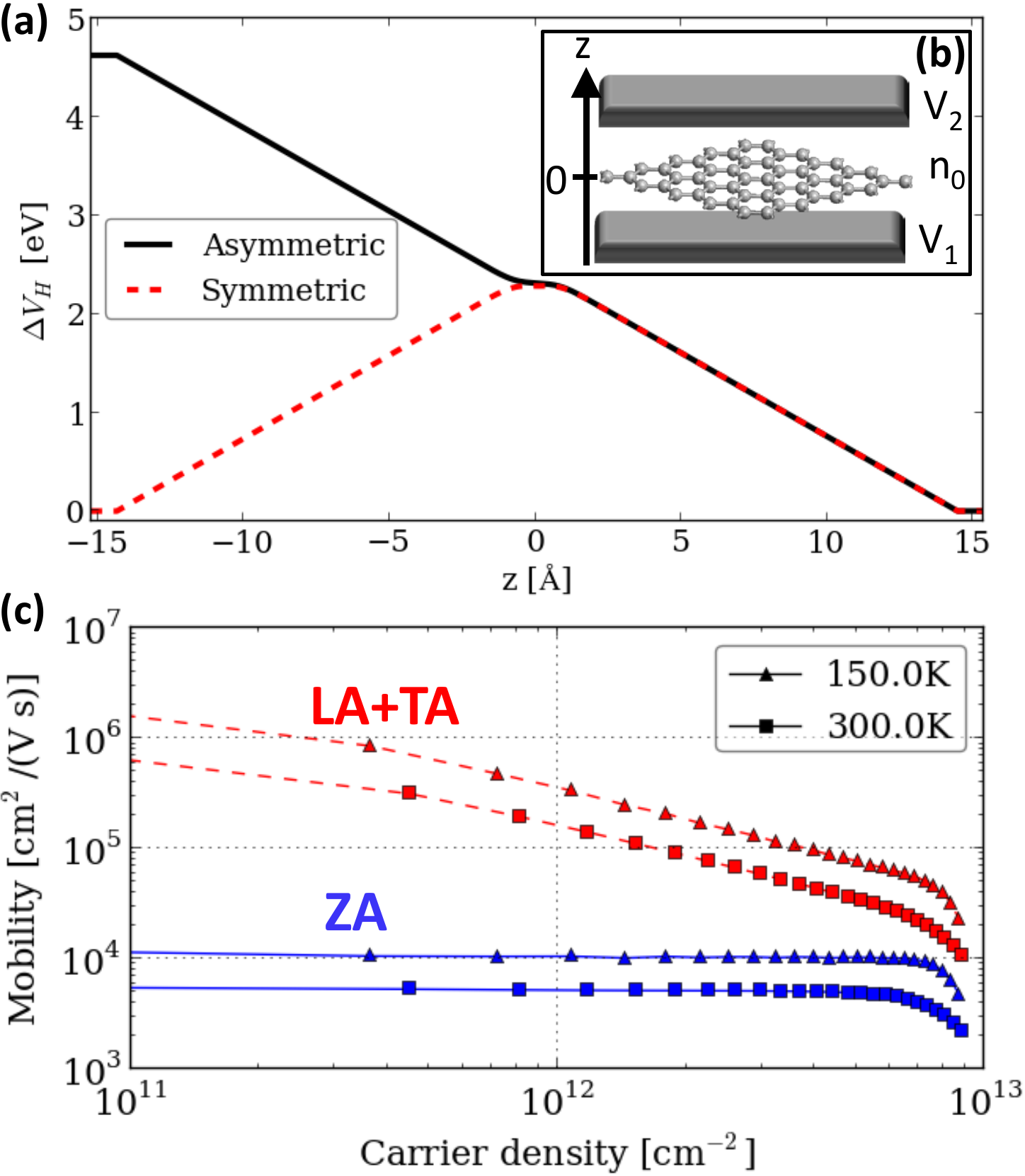}}
	\caption{Electrode stack setup of a graphene device. (a) Potential profile across the graphene device in a symmetric and asymmetric arrangement. (b) Arrangement consisting of graphene and two electrodes. In the symmetric mode the electrodes are kept at the same potential $V_1=V_2$ and a charge is added to the graphene layer. In the asymmetric mode an electric field is generated by a potential difference between the two electrodes $V_1=-V_2$. The mobility is only degraded in the asymmetric mode. (c) Mobility at two different temperatures without (TA+LA) and with ZA scattering in a device with graphene sandwiched between two different dielectrics. The dielectric regions have $\kappa_A=2.5$ and $\kappa_B=22$.
	}
	\label{fig:SystemDoubleGateFields}
\end{figure}
In Fig.~\ref{fig:TemperatureDependence} we show the energy dependence of the momentum relaxation rate for the in-plane $\tau_\text{TA+LA}^{-1}=\tau_\text{TA}^{-1}+ \tau_\text{LA}^{-1}$ and flexural $\tau_\text{ZA}^{-1}$ modes at $T=20$~K and $T=300$~K, and  $n_0=2\times10^{13}$~cm$^{-2}$.
In the vicinity of the Fermi level, the maximum momentum transfer of quasielastic scattering is limited to $q_\text{max}=2k_F$ where $k_F$ is the Fermi wavevector. The corresponding energy, $\hbar \omega_{q_\text{max}}$, defines the BG temperature, $k_B T_{\text{BG}} = \hbar \omega_{q_\text{max}}$, below which short-wavelength phonon scattering is frozen out. For the flexural phonon, $k_BT_{\text{BG}} = b q_\text{max}^2$ is significantly smaller than the one for the linear in-plane phonons, $k_BT_{\text{BG}} = \hbar c_s q_\text{max}$, where $c_s$ is the sound velocity. We obtain $T_{\text{BG}} \approx 57\, \sqrt{\tilde{n}}$~K ($\approx 0.46\, \tilde{n}$~K) for the in-plane (flexural) modes, where $\tilde{n} = n/10^{12}$~cm$^{-2}$~\cite{FN_graphene_parameters}.
At 20~K, the in-plane acoustic phonons are in the BG regime where the reduced phase space available for phonon scattering manifests itself in a pronounced dip in $\tau_\text{TA+LA}^{-1}$ at the Fermi level which is not present at $T=300$~K. The dip is also absent for the field-induced ZA scattering which remains in the equipartition (EP) regime at both temperatures. Based on the low-energy description of the field-induced ZA coupling in Eq.~\eqref{eq:M}, we find
\begin{align}
  \label{eqn:EPregimeZA}
  \frac{1}{\tau_k^{ZA}} & = \frac{v_F D_0^2(V_G)}{2 \pi \rho b^2} 
                            \varepsilon_k^{-2} \frac{k_B T}{q_c} ,
\end{align}
where $q_c$ is the momentum cutoff discussed previously and the linear temperature dependence originates from equipartitioning. In Fig.~\ref{fig:TemperatureDependence}(a), scattering off the ZA mode clearly dominates at both temperatures; it is almost an order of magnitude higher than the scattering off the TA and LA modes at 300~K and even more at 20~K.

The impact on the temperature dependence of the mobility is illustrated in Fig.~\ref{fig:TemperatureDependence}(b) which shows the temperature exponent $\alpha$ of the mobility $\mu \propto T^{-\alpha}$. For in-plane acoustic phonon scattering, $\alpha=1$ in the EP regime and $1 < \alpha \lesssim 5$ in the BG regime~\cite{hwang_acoustic_2008,kaasbjerg_unraveling_2012} in good agreement with the curves for isolated graphene in Fig.~\ref{fig:TemperatureDependence}(b) where field-induced ZA scattering is absent. In the presence of ZA scattering, the mobility acquires the linear temperature scaling, $\alpha \approx 1$, of the momentum relaxation rate in Eq.~\eqref{eqn:EPregimeZA} even at low temperatures. This is a direct consequence of the fact that field-induced ZA scattering remains in the EP regime even at high carrier densities and low temperatures. We note that this differs from two-phonon ZA scattering that shows $\alpha \approx 2(4)$ in the EP (BG) regime.\cite{ochoa_scattering_2012,castro_limits_2010} In the EP regime the field-induced scattering is indistinguishable from in-plane scattering since both scales linearly with $T$. This could explain the high coupling constants experimentally extracted at 300\,K on gated graphene devices.

The modified density dependence of the field-induced ZA scattering dominated mobility in Fig.~\ref{fig:SystemCapacitor}(c), stems from the energy dependence of the momentum relaxation rate in Eq.~\eqref{eqn:EPregimeZA} and the fact that the field-induced coupling $D_0(V_G)$ depends on the carrier density via the gate potential. We therefore have that $\tau_\text{ZA}^{-1} \propto \varepsilon_F^2$, since $D_0(V_G) \sim n_0 \propto \varepsilon_F^2$, whereas $\tau_\text{TA+LA}^{-1}\propto \varepsilon_F$ ($\propto 1/\varepsilon_F^2$) for in-plane acoustic phonon scattering in the EP (BG) regime. For the density dependence of the mobility, $\mu\approx ev_F^2\tau_{k_F}/\varepsilon_F$, this implies a change from $\mu \sim 1/n_0$ ($\sqrt{n_0}$) in the EP (BG) regime to $\mu\sim n_0^{-3/2}$ when field-induced ZA scattering dominates in-plane acoustic phonon scattering. This explains the opposite trends with $n_0$ observed in Fig.~\ref{fig:SystemCapacitor}(c).

Finally, we point out that it is possible to control the symmetry-breaking field, and thus the field-induced coupling to the flexural phonons, and the carrier density independently in experiments. In Fig.~\ref{fig:SystemDoubleGateFields}(a), we consider a setup with two gate electrodes which can be biased independently. In the asymmetric gate configuration an electric field is generated by a potential difference between the two electrodes $V_1=-V_2$. This results in a potential that breaks the $\sigma_h$ symmetry and a mobility reduction corresponding to that in Fig.~\ref{fig:SystemCapacitor}(b). Alternatively, one can introduce charge carriers in graphene with a symmetric gate configuration where both electrodes are kept at the same potential $V_1=V_2$. In this case, the potential preserves the $\sigma_h$ symmetry and no field-induced scattering occurs. Even in the absence of a gate potential we may introduce the flexural scattering due to different dielectrics on each side of graphene. This is illustrated in Fig.~\ref{fig:SystemDoubleGateFields}(c), where the effect originates solely from the dielectric environment.

We have shown that one-phonon flexural phonon scattering can be activated in graphene devices depending on the symmetry of the electrostatic and dielectric environment.
This field-induced flexural phonon scattering was demonstrated to have a detrimental impact on the performance of graphene.
The mechanism is indistinguishable from in-plane scattering at room temperature and could hereby explain the high coupling constants consistently needed to explain experiments on gated graphene devices.
In addition, we showed that this scattering modifies temperature and density scaling of the mobility, which allows for its experimental verification. Paradoxically, better sample quality may show worse performance due to a lower cutoff of the long-wavelength scattering. Protecting the planar mirror symmetry is therefore of utmost importance to fully exploit the unique transport properties of graphene.

\newpage
\begin{acknowledgments}
The authors acknowledge support from Innovation Fund Denmark(Grant No. 79-2013-1), 
and CNG is sponsored by the Danish National Research Foundation, project No. DNRF103.
\end{acknowledgments}

\bibliography{GrapheneInAField}

\begin{thebibliography}{47}
\bibitem [{FN_({\natexlab{a}})}]{FN_settings}%
  \BibitemOpen
  \href@noop \ \bibinfo {note} {We use ATK DFT
  with PBE-GGA exchange-correlation, Double-Zeta-Polarized
  basis-set and $51\times51$ in-plane $k$-points. At finite doping, a charge is added to the
  graphene layer ($z=0$~\AA) and the Poisson equation is solved with Dirichlet (Neumann)
  boundary condition at the gate electrode, $z=-15$~\AA (vacuum boundary, $z=15$~\AA).
}\BibitemShut
  {Stop}%
\bibitem [{FN_({\natexlab{b}})}]{FN_PHandEPH}%
  \BibitemOpen
  \href@noop \ \bibinfo {note} {The geometry was relaxed
  until all forces were smaller than $0.001~\mathrm{eV/\AA}$. The el-ph
  coupling and phonon dispersion were obtained from a 11x11 supercell
  calculation, with the el-ph evaluated at each gate-value.}\BibitemShut
  {Stop}%
\end{thebibliography}

\end{document}